# Highly-asymmetric soliton complexes in parabolic optical lattices

Yaroslav V. Kartashov, Victor V. Vysloukh, and Lluis Torner

*ICFO-Institut de Ciencies Fotoniques, and Universitat Politecnica de Catalunya,*

*Mediterranean Technology Park, 08860 Castelldefels (Barcelona), Spain*

We introduce multipole soliton complexes in optical lattices induced by nondiffracting parabolic beams. Despite the symmetry-breaking dictated by the curvature of the lattice channels, we find that complex, asymmetric higher-order states can be stable. The unique topology of parabolic lattices affords new types of soliton motion: single solitons launched into the lattice with nonzero transverse momentum perform periodic oscillations along parabolic paths.

The propagation of light in transversely inhomogeneous refractive index landscapes gives rise to a plethora of phenomena that cannot occur in uniform media. The symmetry of the refractive index profile is the crucial ingredient that sets the properties of nonlinear light excitations. One way to imprint such structures is optical induction [1-4]. Plane waves are the simplest examples of nondiffracting beams that can be used for lattice induction. Nontrivial nondiffracting beams are associated with solutions of Helmholtz equation in coordinate systems that allow its separability into the transverse and longitudinal parts. Bessel, Mathieu, and parabolic beams are fundamental solutions in circular, elliptical, and parabolic coordinates, respectively. The symmetries of such beams can impart important new properties for solitons in the corresponding optical lattices. Radially symmetric stable solitons, vortices, and rotary soliton motion were demonstrated in Bessel lattices [5-8]. The topological transformation between Bessel and periodic lattices, accompanied by a drastic modification of soliton properties, is one of the signatures of Mathieu lattices [9]. Nonlinear Schrödinger models on curved one-dimensional manifolds predict curvature induced symmetry breaking [10]. The properties of localized excitations in a single parabolic channel were analyzed [11], but the salient features of solitons supported by lattices produced by parabolic beams [12] with multiple curved channels remain unexplored to date.

In this Letter we introduce the unique properties exhibited by solitons in lattices produced by parabolic nondiffracting beams. Parabolic lattices feature nonzero curvature of its channels that results in asymmetric shapes of higher-order solitons. We predict that despite such symmetry breaking, complex higher-order states can be stable. We show that the specific topology of parabolic lattices affords oscillatory-type soliton motion.

We consider beam propagating along the $\xi$-axis of a photorefractive medium with focusing nonlinearity in a lattice induced by nondiffracting beam. The beam propagation is described by the nonlinear Schrödinger equation for the field amplitude $q$ [1-4]:

$$i\frac{\partial q}{\partial \xi} = -\frac{1}{2}\left(\frac{\partial^2 q}{\partial \eta^2} + \frac{\partial^2 q}{\partial \zeta^2}\right) - Eq\frac{|q|^2 + pR}{1 + |q|^2 + pR}, \quad (1)$$

where the transverse $\eta, \zeta$ and longitudinal $\xi$ coordinates are normalized to the beam width and to the diffraction length, respectively; the parameter $E$ stands for the static dc electric field; $p$ is the lattice depth; the function $R(\eta,\zeta)$ describes the lattice profile. We let the lattice feature the intensity of a nondiffracting parabolic beam $R \sim |q_{\rm nd}|^2$, where

$$q_{\rm nd} = \exp(-ib_{\rm lin}\xi)\int_{-\pi}^{\pi} A(\phi)\exp[i(2b_{\rm lin})^{1/2}(\eta\cos\phi + \zeta\sin\phi)]d\phi. \quad (2)$$

Here $A(\phi)=0$, for $\phi\in[-\pi,0)$; and $A(\phi)=(\pi|\sin\phi|)^{-1/2}\exp(ia\ln|\tan\phi/2|)$, for $\phi\in[0,\pi)$; is the angular spectrum of the parabolic beam. Such beams are formed by the illumination of an annular slit with properly selected angular transmission function [12]. The function $R$ is normalized so that $\max R = 1$. Parabolic lattice features multiple curved channels, with the first of them being the most pronounced [Fig. 1(a)]. The channel density gradually increases, while the lattice amplitude decreases towards its periphery. The parameter $a$ in Eq. (2) determines the curvature of the lattice channels. Increasing $a$ causes a decrease of the curvature. Changing the parameter $b_{\rm lin}$ results in a simple rescaling. Here we set the representative values $a = 2$, $b_{\rm lin} = 4$, and $E = 10$ in Eq. (1), which corresponds to typical experimental conditions.

We look for soliton solutions of Eq. (1) in the form $q(\eta,\zeta,\xi) = w(\eta,\zeta)\exp(ib\xi)$, where $b$ is the propagation constant. Upon linear stability analysis the perturbed solutions were searched in the form $(w + u + iv)\exp(ib\xi)$, where perturbation $u + iv$ can grow with a complex rate $\delta$ upon propagation. Linearization of Eq. (1) leads to eigenvalue problem (see [13] for details) that we solved numerically. Parabolic lattices support a variety of linear guided modes residing in the first channel, both fundamental and multipole ones (i.e. those featuring $\pi$ phase jumps between bright spots). The linear spectrum is important for analysis of localized nonlinear states. Figure 2(e) illustrates the dependence of cutoffs of linear propagation constants on the mode number. Notice that linear eigenstates residing in the first channel of the lattice are somewhat similar to modes of parabolically curved channel [11]. Soliton families bifurcate from linear guided modes (see Fig. 1 for examples of soliton profiles). Fundamental solitons are strongly elliptical near the lower cutoff $b_{\text{low}}$, while with increase of $b$, accompanied by a monotonic growth of the energy flow $U = \int_{-\infty}^{\infty} |q|^2 \, d\eta d\zeta$ [Fig. 2(a)], solitons become almost circularly symmetric. Besides the lower cutoff $b_{\text{low}}$, where $U \to 0$, there also exists an upper cutoff $b_{\text{upp}} < E$ where, however, the energy flow does not diverge as in uniform saturable media; rather, it remains finite. The upper cutoff appears because in the regime of strong saturation soliton tend to expand in the transverse direction. In bulk media such expansion is uniform. However, in the lattice solitons expansion is enhanced in high refractive-index areas, which is equivalent to repulsion from the interface between modulated and uniform regions. An analogous phenomenon was addressed in Ref. [13]. Increasing the peak amplitude results in enhanced soliton mobility, so that at a critical power the lattice can no longer compensate the repulsive force. Solitons in the upper cutoff are well localized. The lower cutoff monotonically increases with $p$, while the upper cutoff decreases [Fig. 2(b)]. The maximal energy flow of fundamental soliton gets smaller in deeper lattices. The soliton ellipticity is mostly pronounced at $b \to b_{\text{low}}$. Ellipticity defined as the ratio of soliton widths along the $\zeta$ and $\eta$ directions in low-power limit, first increases with $p$ [Fig. 2(c)], but then starts decreasing since the lattice itself washes out at high $p$. The soliton width first decreases with $b$, but then starts increasing due to saturation. By solving the eigenvalue problem for perturbations, we found that fundamental solitons are stable in the entire domain of their existence.

Besides fundamental solitons we found a variety of higher-order solitons supported by parabolic lattices. The simplest of them (dipole) consists of two out-of-phase bright spots.

Such solitons are strongly asymmetric [Fig. 1(d)]. As in the case of fundamental solitons an abrupt shift into the lattice depth occurs when the energy flow of the dipoles becomes too high. Because of the additional repulsion acting between the poles, such a shift occurs at even smaller energy flows than in the case of fundamental solitons. As a result, the upper cutoff for dipoles is smaller than that for fundamental solitons [Fig. 2(d)]; also, it gradually increases with $p$. For all multipole solitons $b_{\text{low}}$ is smaller than that for fundamental ones. The important result of this Letter is that despite the symmetry breaking, complex multipole states can be stable in largest part of their existence domain. Only close to the upper cutoff, when strong saturation results in reduction of lattice effects, small perturbations may result in remarkable oscillations of the spots forming multipoles. Such oscillations are connected with so-called neutral eigenmodes (see Ref. [14]) that are excited by the input noise. At low and moderate power levels all dipoles, triple-mode solitons [Fig. 1(e)] and higher-order multipoles are stable, also in the presence of strong input noise, as shown in Fig. 3. This result is important, since most of nonlinear states in optics featuring significantly curved shapes tend to be dynamically unstable. We also found solutions residing in different channels of parabolic lattice [see Fig. 1(f) for a triple-mode soliton featuring two spots in the first channel and one spot in the second channel]. The second lattice channel also supports fundamental solitons, but their existence domain is rather narrow [see Fig. 2(f)].

We also found that the geometry of parabolic lattice affords new types of beam dynamics. To illustrate it, we imposed an initial phase tilt $\exp(i\alpha\zeta)$ on the fundamental soliton. At $\alpha > 0$ such solitons start oscillating in the widest lattice channel upon propagation, so that the integral soliton center moves along a trajectory which is close to a parabolic one [Fig. 4(c)]. The maximal soliton center displacement along $\zeta$ axis, $d_\zeta$, monotonically increases with $\alpha$, while $\xi$-period of oscillations, $d_\xi$, grows with $\alpha$ quite rapidly [Figs. 4(a) and 4(b)]. For large $\alpha$ values one clearly observes anharmonicity of oscillations.

Summarizing, we introduced the unique properties of multipole solitons in lattices that can be created by an important class of nondiffracting beams which feature parabolic shapes. We found that despite the symmetry breaking dictated by the curvature of the parabolic lattice channels, such multipole solitons can be stable.

# References with titles

# References without titles

# Figure captions

Figure 1. Parabolic lattice corresponding to $a=2$ and $b_{\text{lin}}=4$ (a) and field modulus distributions for fundamental soliton (b) and (c), dipole soliton (d), and triple-mode solitons residing in (e) first lattice channel and (f) first and second lattice channels. Propagation constant values are indicated on the plots. In all cases $p=4$.

Figure 2. (a) Energy flow versus $b$ for fundamental solitons. Circles correspond to profiles shown in Figs. 1(b) and 1(c). Cutoffs (b) and ellipticity in low-power limit (c) versus $p$ for fundamental solitons. (d) Cutoffs versus $p$ for dipole solitons. (e) Cutoffs for linear guided modes at $p=4$. (d) Energy flow versus $b$ for soliton from second lattice channel.

Figure 3. Stable propagation of (a) dipole soliton with $b=7.2$, (b) triple-mode soliton with $b=6$, and (c) five-hump soliton with $b=6$ in the presence of noise with variance $\sigma^2_{\text{noise}}=0.01$ at $p=4$. Field modulus distributions are shown at different distances.

Figure 4. Maximal displacement along $\zeta$-axis (a) and period of soliton oscillations (b) versus $\alpha$. (c) Snapshot images showing soliton oscillations at $\alpha=1$. Images are taken at $\xi=0$, $\xi=d_\xi/4$, and $\xi=3d_\xi/4$. White line indicates trajectory of soliton center. In all cases the input soliton was taken at $b=6.5$ and $p=2$.

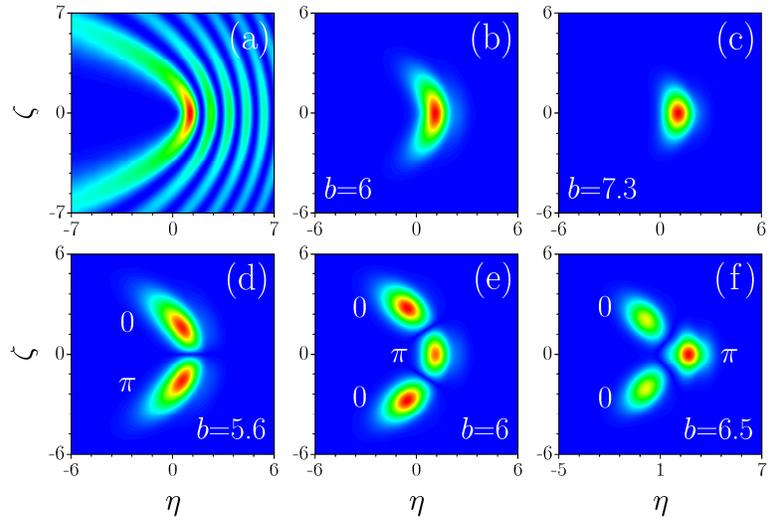

Figure 1. Parabolic lattice corresponding to $a=2$ and $b_{\mathrm{lin}}=4$ (a) and field modulus distributions for fundamental soliton (b) and (c), dipole soliton (d), and triple-mode solitons residing in (e) first lattice channel and (f) first and second lattice channels. Propagation constant values are indicated on the plots. In all cases $p=4$.

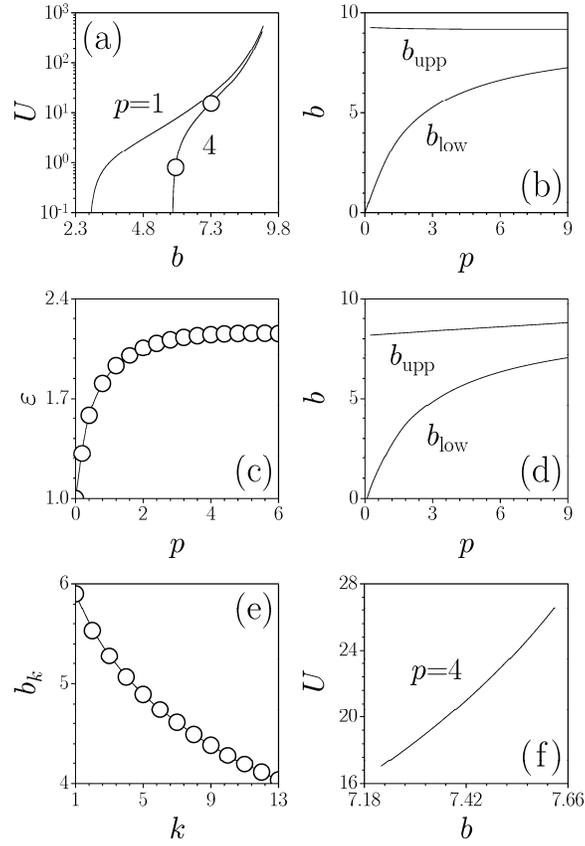

Figure 2. (a) Energy flow versus $b$ for fundamental solitons. Circles correspond to profiles shown in Figs. 1(b) and 1(c). Cutoffs (b) and ellipticity in low-power limit (c) versus $p$ for fundamental solitons. (d) Cutoffs versus $p$ for dipole solitons. (e) Cutoffs for linear guided modes at $p=4$. (d) Energy flow versus $b$ for soliton from second lattice channel.

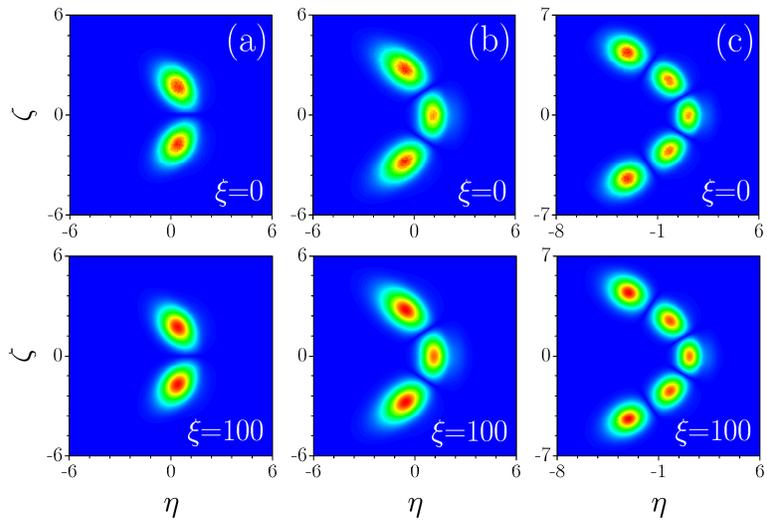

Figure 3. Stable propagation of (a) dipole soliton with $b = 7.2$, (b) triple-mode soliton with $b = 6$, and (c) five-hump soliton with $b = 6$ in the presence of noise with variance $\sigma_{\text{noise}}^2 = 0.01$ at $p = 4$. Field modulus distributions are shown at different distances.

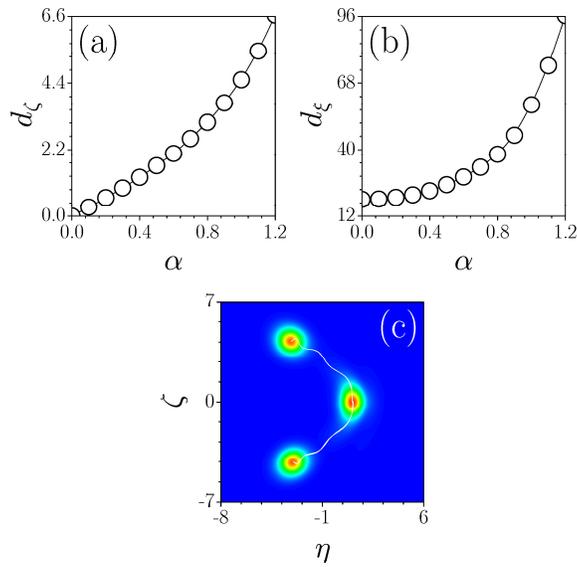

Figure 4. Maximal displacement along $\zeta$-axis (a) and period of soliton oscillations (b) versus $\alpha$. (c) Snapshot images showing soliton oscillations at $\alpha = 1$. Images are taken at $\xi = 0$, $\xi = d_\xi/4$, and $\xi = 3d_\xi/4$. White line indicates trajectory of soliton center. In all cases the input soliton was taken at $b = 6.5$ and $p = 2$.

12